\documentclass[preprint2]{aastex}

\usepackage{amsmath}

\usepackage{color}

\begin{document}

\title{Non-parametric study of the evolution of the cosmological equation of state with SNeIa, BAO and high redshift GRBs}
\author{S. Postnikov}
\affil{Indiana University}
\affil{Nuclear Theory Center, Bloomington IN, USA}
\email{spostnik@indiana.edu}
\and
\author{M. G. Dainotti}
\affil{Stanford University}
\affil{Physics Department, Via Pueblo Mall, 382, Stanford CA, USA;\newline
Jagiellonian University, Golebia, 24, Krakow, Poland}
\email{mdainott@stanford.edu;dainotti@oa.uj.edu.pl}
\author{X. Hernandez}
\affil{Instituto de Astronom\'{\i}a, Universidad Nacional Aut\'onoma de M\'exico, M\'exico D.F. 04510, M\'exico}
\email{xavier@astros.unam.mx}
\and
\author{S. Capozziello} 
\affil{Dipartimento di Fisica,  Universit\'{a} di Napoli "Federico II", Compl. Univ. di Monte S. Angelo, Edificio G, 
Via Cinthia, I-80126 - Napoli, Italy and INFN Sez. di Napoli, Italy}
\email{capozziello@na.infn.it}

\begin{abstract}

We study the dark energy equation of state as a function of redshift
in a non-parametric way, without imposing any {\it a priori} $w(z)$ (ratio of pressure over energy density) functional form. As a 
check of the method, we test our scheme through the use of synthetic data sets produced from different input cosmological models 
which have the same relative errors and redshift distribution as the real data. Using the luminosity-time $L_{X}-T_{a}$ 
correlation for GRB X-ray afterglows (the Dainotti et al. correlation), we are able to utilize GRB sample from the {\it Swift} 
satellite as probes of the expansion history of the Universe out to $z \approx  10$. Within the assumption of a flat FLRW universe 
and combining SNeIa 
data with BAO constraints, the resulting maximum likelihood solutions are close to a constant $w=-1$. If 
one imposes the restriction of a constant $w$, we obtain $w=-0.99 \pm 0.06$ (consistent
with a cosmological constant) with the present day Hubble constant as $H_{0}=70.0 \pm 0.6$
${\rm km} \, {\rm s}^{-1} {\rm Mpc}^{-1}$ and density parameter as $\Omega_{\Lambda 0}=0.723 \pm 0.025$, while 
non-parametric $w(z)$ solutions give us a probability map which is centred at $H_{0}=70.04 \pm 1$
${\rm km} \, {\rm s}^{-1} {\rm Mpc}^{-1}$ and $\Omega_{\Lambda 0}=0.724 \pm 0.03$. Our chosen GRB data sample 
with full correlation matrix allows us to estimate the amount, as well as quality (errors) of data, needed to constrain 
$w(z)$ in the redshift range extending an order of magnitude in beyond the farthest SNeIa measured.

\end{abstract}


\section{Introduction}

Recent advances in precision cosmology and astronomical observations have yielded large amounts of data against which 
cosmological models can be calibrated and tested. Among them are supernovae type Ia (SNeIa) catalogues, galaxy surveys, baryonic 
acoustic oscillation (BAO) measurements e.g. \citep{Tegmark2006}, temperature fluctuations of the cosmic microwave background (CMB), 
observational Hubble data from differential ages of galaxies (OHD) e.g. \citep{Hz_2012, HzI_2009, HzII_2009} and ages of Globular 
Clusters (GC) e.g. \citep{Mackey2002tn,Mackey2002un,Mackey2002jm}.

In principle, taking the above as statistically independent singular event measurements, the expansion history of the universe and 
its temporal evolution, can be inferred. Most inferences are observationally limited to the region $z<2$, at best, with reliable low error 
observations gradually disappearing after $z=1.5$. 

Given the much larger redshift range over which gamma ray bursts (GRBs) 
can be observed extending out to $z\approx 10$  e.g.\citep{Izzo2009} it has long been tempting to include GRBs as cosmological probes, 
extending the redshift range by almost an order of magnitude further than available SNeIa.
To this end it is necessary to find scaling relations among GRB observables. 

A crucial breakthrough in this field has been the observation of GRBs by the \textit{Swift} satellite which provides a rapid follow-up of 
the afterglows in several wavelengths with better coverage than previous missions. {\it Swift} revealed a more complex behavior of the 
lightcurves, rather different from the broken power\,-\,laws assumed in the past. The lightcurves observed by Swift identify two, 
three and sometimes even more regions in the afterglows. The second segment, a temporally extended phase of close to constant
luminosity, is called the plateau region.
Here we make use of a proposal by some of us, who in \citet{Dainotti2008,Dainotti2010,Dainotti2011a,Dainotti2011b} discovered that a 
tight correlation exists in GRB light curves between the X-ray luminosity at the end of the plateau phase, $L_X$, and the rest frame temporal 
duration of this phase, $T^{*}_a=T^{obs}_a/(1+z)$. 

In this paper we develop a new fully Bayesian (reduced to maximum likelihood) methodology which we apply to the problem of inferring 
the evolution 
of the dark energy equation of state, $w(z)$, firmly based upon the best existing $z<2$ observations, and traced for the first time 
out to $z\approx 9$ through the use of a carefully selected subsample of $68$ long GRBs. An advantage of our statistical approach is that 
no {\it a priori} functional form for $w(z)$ is imposed, the method proceeds from a large number of randomly chosen $w(z)$ functions, 
which are refined to yield an optimum maximum likelihood answer, together with corresponding confidence interval bands. 

The non-parametric 
reconstruction of the dark energy equation of state (DE EoS) has been recently addressed by several authors e.g. 
~\citep{Su2012, Barboza2009, Holsclaw2010, Holsclaw2011, Adak2011, Barboza2011}. Sophisticated Bayesian analysis was done by 
Vazquez et al.~\citep{Vazquez2012} to reconstruct the dark energy equation of state, who find a mild indication of time-dependence within 
the low redshift range available to non-GRB observations. Our sequential Bayesian analysis method arises from similar work being done in 
reconstructing the equation of state of dense nuclear matter from observational properties of neutron stars. The method needs no binning 
or predetermined $w(z)$ proposals, and so, uses fully all the information available in the data, which are then the only factor driving 
the solution. The use of the recently established correlation for a sub-sample of GRBs then in principle allows for a reconstruction of 
the evolution history of the dark energy equation of state out to an unprecedented redshift of 8.2. Unfortunately, the current
number of high redshift events, and their large error bars, permit only very crude $w(z)$ inferences at high redshift.

Still, we obtain a maximum likelihood solution for the optimum $w(z)$ model, remaining within the assumptions of spatial flatness, 
standard general relativity and Friedmann-Lema\^{i}tre -Robertson-Walker \\
(FLRW) models yielding highly accurate determinations for
$H_{0}$ and $\Omega_{\Lambda 0}$. Although the resulting confidence interval bands allow for some minor variations for the low redshift 
$w(z)$ of $\pm 0.3$, the result we obtain is consistent with a classical cosmological constant $w(z)=-1$. We note that our results, 
as is always the case, do not represent an absolute confirmation of the assumptions made, in this case a classical GR universe with 
spatial flatness, which would have to be validated or excluded using a more extended approach which considers also other cosmological 
models as those related to extended theories of gravity \citep{CapFra98, CapdLa11}. 

Section(2) discusses the use of the $L_{X}-T_{a}$ correlation for a sub-set of GRBs, section (3) gives an introduction to the basic 
sequential Bayesian approach, and describes the cosmological model under which we will be working, a 
Friedmann-Lema\^{i}tre -Robertson-Walker (FLRW)  model. The maximum likelihood $w(z)$ reconstruction methodology applied to the available 
data sets is developed in section (4), where it is also tested through the use of synthetic data samples constructed for a series of
test input cosmological models. In this same section we detail the construction of a distance ladder, in order to use high redshift GRBs, 
and give also the treatment of correlations among GRB data points. Section (5) gives our results using the real data samples, and 
section (6) presents our conclusions.

\section{GRBs as distance estimators}

Despite the extraordinary redshift range over which GRBs are observed, their use as cosmological references has been hindered by the fact 
that they are not inherently standard candles of fixed intrinsic luminosity. Numerous efforts have been made over the past years, looking 
for correlations between the parameters of GRB light curves which might serve to turn GRBs into cosmological measurement tools. A number 
of published correlations e.g. $E_{iso}$\,-\,$E_{peak}$ \cite{Lloyd2000,amati09}, $E_\gamma$\,-\,$E_{peak}$ \cite{G04,Ghirlanda06}, 
$L$\,-\,$E_{peak}$ \cite{S03,Yonekotu04}, $L$\,-\,$V$ \cite{FRR00,R01} and other proposed luminosity indicators \cite{N000,liza05,LZ06} 
however, suffer from the problem of large data scatter e.g. \cite{Butler2009,Yu09}. {\color{black}Also, the probable impact of detector thresholds on 
cosmological standard candle calibrations \citep{Shahmoradi2011ck} is an issue which has been debated e.g. \cite{Cabrera2007}, and 
should be taken into account.}

The underlying problem of the scatter in the above correlations is that it is larger than the spread expected from the redshift
dependence alone. GRBs can be seen over a large fraction of the visible Universe, up to z=9.4 \cite{Cucchiara2011}. The luminosity spread 
due to, exclusively the luminosity distance squared dependence, gives for the limiting redshifts, a factor of 
$D_L^2(9.4)/D_L^2(0.085)=6.4 \times 10^{4}$, while the actual spread in luminosity is of 8 orders of magnitude, from 
$10^{46}$ to $10^{54}$ ergs/s. It is not clear what is responsible for such a large dynamical range. Moreover, the correlations listed 
above are affected by selection biases especially at high redshifts where only the more luminous events are detected (Malmquist effect).
 Lastly, in the application of GRBs to cosmology, possible logical circularity problems should be carefully taken into account and 
explicitly excluded e.g. \citep{Petrosian2009}. 

To overcome this problem, \citet{Dainotti2013ApJ} demonstrated through the Efron and Petrosian method \citep{Efron1992} that the Dainotti 
correlation is not an apparent correlation, but is due to the intrinsic properties to GRBs, at a 12 $\sigma$ level, with an intrinsic 
slope of $b_{intrinsic}=-1.07_{-0.14}^{+0.09}$, see eq(1) below. This is a very relevant finding since the Dainotti et al. correlation in 
Eq. \ref{feq} is not only a useful tool for the testing of theoretical GRB physical models e.g. 
\citep{Cannizzo09,Cannizzo11,Dall'Osso,Bernardini2011,Rowlinson2010,Yamazaki09,Ghisellini2008}, but has also been applied to cosmology 
in e.g. \cite{Cardone09,Cardone2010}. Moreover, Dainotti et al. 2013b identified a caveat on the use of non intrinsic correlations to
constraining cosmological parameters, by showing how systematics can lead to errors in the evaluation of the cosmological parameters 
\citep{2013MNRAS43682D}.

This correlation can be stated as:
\begin{equation}
\lg L_X = \lg a + b \lg T^*_{a}.
\label{feq}
\end{equation}
The normalization $a$ and the slope parameter $b$ are constants obtained through the D'Agostini fitting method \citep{Dago05}. 
We determine the power law slope $b$ in Eq. \ref{feq} after properly correcting for possible data selection due to instrumental 
threshold effects. This step is necessary to make the Dainotti correlation a useful distance estimator and an effective cosmological tool.
Notice crucially that the correlations present in the literature listed above, have as one variable $E_{iso}$ and because of that
they suffer from a double truncation due to detection selection thresholds \citep{Lloyd1999ApJ}. Thus, these correlations bring with 
themselves a dual selection bias problem both at low energy and high energy. In the Dainotti et al. correlation, by involving time, 
which does not depend on the detector threshold, and depending on the luminosity of the plateau $L_{X}$, the only problem is the 
detection threshold for faint plateau cases. Thus, selection biases are significantly reduced in comparison to other correlations.

Regarding the sample analyzed, it is constituted by all GRB X-ray afterglows with known redshifts detected by {\it Swift} from January 2005 
up to May 2011, for which the light curves include early X-ray Telescope data and therefore can be fitted using Willingale's phenomenological 
model \citep{W07}. We used the redshifts available in the literature \citep{X09} and in the Circulars Notice arxive (GCN), after excluding 
all GRBs with non-spectroscopic redshifts.

In previous papers \citep{Dainotti2008,Dainotti2010,W07} the {\it Swift} Burst Alert Telescope (BAT)+ X-Ray Telescope (XRT) light curves 
of GRBs were fitted with a two component model assuming that the rise time of the afterglow, $t_a$, started at the time of the beginning 
of the decay phase of the prompt emission, $T_p$, namely $t_a=T_p$. Here we search for an independent measure of the above parameters 
of the afterglow, thus leaving $t_a$ as a free parameter. In the majority of cases we have $t_a \geq 0$. We have created a semi-automatized 
analyzer with the software Mathematica 9 which allows a computation of the best fit parameters \footnote{ASCII tables with all the quantities 
needed for the analysis and the MATHEMATICA codes used are available on request.}. 
  
The source rest-frame luminosity in the {\it Swift} XRT bandpass, $(E_{min}, E_{max})=(0.3,10)$ keV, is

\begin{equation}
\begin{split}
L_X & (E_{min},E_{max},t)= \\
& 4 \pi d_L^2(z) \, F_X (E_{min},E_{max},t) \cdot \textit{K} ,
\end{split}
\label{eq: lx}
\end{equation}
where $d_L(z)$ is the GRB luminosity distance for the redshift $z$, and a flat cosmological model with any chosen
$\Omega_M$, $w(z)$ and $h$ parameters. $F_X$ is the measured X-ray energy flux in ${\rm erg \, cm^{-2}  s^{-1}}$ and  \textit{K} is 
the \textit{K}-correction for cosmic expansion. Details on the fitting procedure and the computation of the power law spectrum can be 
found in \cite{Evans2009,Dainotti2010}.

The complete sample of GRBs analysed contains 101 events, covering the redshift range $0.033 \leq z \leq 9.4$. In our analysis, we
 take a subsample of GRBs within the SNe Ia overlap redshift range of $z<1.4$, and use them to derive the best fit parameters 
$a$ and $b$ through the D'Agostini method for the LT correlation. We found for this subsample $b_{obs}=-1.51_{-0.27}^{+0.26}$, however, we 
are aware from Dainotti et al 2013a  that the steepening of the slope is an effect of time evolution, therefore we use in the analysis 
the intrinsic slope b. This intrinsic
 slope has been evaluated assuming an underling flat cosmology for the Universe. The small number of GRBs in the overlap region prevents us 
from repeating the EP test. However, we also point out that in Dainotti et al. 2013b we have demonstrated that the calibration of the
correlation with the present data set is independent of the underling cosmology within the error ranges of the parameters themselves.
Further, within the overlap region, SNe Ia and a host of other independent cosmological tracers robustly fix the low z cosmological model.
This make our assumption of using $b_{int}$ robust for our purposes. One can then use the same distance luminosity we have obtained from the 
SNe Ia sample with these parameters calculated for $z<1.4$. Thus, we assume the cosmology inferred from the SNe Ia is valid in the low redshift
range, and we then use the GRB correlation to calculate distance luminosities for the high redshift sample. In this way, the circularity 
problem is eliminated, because no assumptions have been made for the high redshift Universe. As will be shown explicitly in section (5), 
we can confirm that the  sample selection is not biasing the results obtained, while it still covers an ample redshift range of $0.49 
\leq z \leq 8.2$. 

\section{Sequential Bayesian analysis and cosmological assumptions}

In this section we outline the basic probabilistic and cosmological assumptions of our work. Bayes' theorem 
~\citep{Bayes1763}, an application of the formula of conditional probability to data sets as arising from a given underling model, 
can be formally written in the following way:
\begin{equation}
	\label{Bayesian_th}
	P(H_y|D) = P(D|H_y) P(H_y)/P(D),
\end{equation}
where $H_y$ is the hypothesis (model) to be tested against a given data set $D$, and $P(H_y)$ is the probability of the hypothesis being 
valid, in absence of data, any prior one might want to introduce. $P(D|H_y)$ is the probability of obtaining the data set $D$ assuming 
$H_y$ to be valid, the likelihood function, $P(H_y|D)$ is probability of the validity of $H_y$ given the occurrence of the data $D$ 
(the posterior) and $P(D)=\sum_{H_y}P(D|H_y)P(H_y)$ the relevant normalization. The key in Bayesian analysis is the introduction of an 
objective and insightful quantitative assignment of probability, in order to take maximum advantage of all the information available in 
observations, mainly central values and confidence intervals of measurements. Our likelihood assignment will be explained in the following 
section.
 
If more statistically independent data become available, the theorem can be applied sequentially to adjust the probability of the 
hypothesis. A sequential formulation has the advantage of being an adaptive approach in analyzing data, consequentially optimizing 
computational time. The approach is highly flexible with respect to handling new data, which enter as refinements on the previous 
solution, with no need of redoing the full analysis. 

The sequential form of Bayes' theorem becomes:
\begin{eqnarray}
	\label{Bayesian_seq}
	P(H_y|[D_i,D_{i-1},...]) \propto && \nonumber \\ 
	P([D_i,D_{i-1},...]|H_y) P(H_y|[D_{i-1},D_{i-2},...]) &&,
\end{eqnarray}
every new data corrects and evolves the probability of the hypothesis to be valid, with previous results effectively becoming a prior on 
the likelihood assignment associated to the new extended data sample. The theorem can thus be thought of as a quantitative formulation 
of Occam's razor.

\begin{figure}[t]
\vspace{-20pt}
	\includegraphics[width=7.5cm]{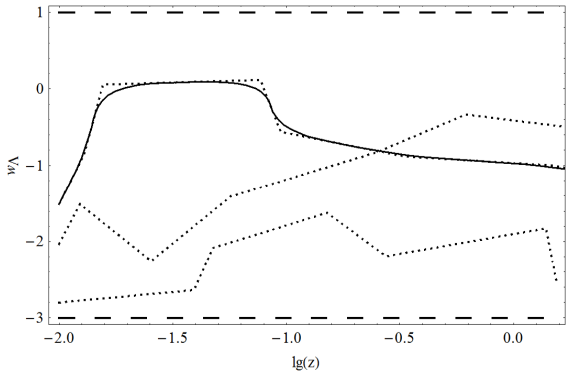}
\vspace{-20pt}
\caption{Examples of randomly generated equations of state inside the chosen boundaries (long dashed lines). Dotted lines represent 
curves generated by choosing a random initial point, slope and line segment length. The solid line is the result of applying the Gaussian 
smoothing filter from Eq. (\ref{gaus1}) with $l=4$.}
\label{exmplDisEOS}
\end{figure}

In this work the hypothesis $H_y$ refers to a particular $w(z)$ function proposed, together with the choice of
two cosmological parameters: the present Hubble constant $H_0$ and the present dark energy density parameter $\Omega_{\Lambda0}$.
We retain the assumption of isotropy for the cosmological model, impose reasonable bounds on the dark energy equation of state 
(see Fig. \ref{exmplDisEOS}) and assume also a constant value for $w(z)$ for the very low $z<0.01$ redshift range. We shall be considering 
a very large number of very general randomly chosen $w(z)$ models restricted only to be continuous in their value, and in their first 
derivative, no violent or rapid transitions are considered. 

To begin, we take no initial priors, i.e., each random EoS is assumed to be equally probable with respect to the others:
\begin{equation}
	\label{init_PH}
P(H_y)=1/N_H,
\end{equation}
where $N_H$ is the total number of generated curves. Each curve is labeled by an index $n$ and receives initial likelihood 
$P(D_1|H_{yn})$ calculated from the analysis of the first set of data with its errors included (represented by $D_1$). As more 
statistically independent data sets are added to the method (represented by $D_i$) the analysis is repeated for new data by the 
sequential multiplication prescribed in Eq. (\ref{Bayesian_seq}). As the number of data sets included increases, the recovered 
$w(z)$ curves with higher relative probabilities will begin to narrow into a band about the optimal $w(z)$. This approach allows 
us to consider independent data sets, and to check the different restrictions each imposes onto the best recovered $w(z)$ model 
and its confidence bands, as they are sequentially added to the full data considered; SN-Ia samples having a low redshift coverage, 
BAO restrictions with no redshift resolution but high local constraining power, and the high redshift GRB sample.

We introduce the standard assumptions of 
spatial flatness, a negligible relativistic component and pressureless matter until the end, in order to have a framework of a very general 
applicability ready. {\color{black}In the case of a redshift independent effective EoS, the
luminosity distance $d_L(z)$ and magnitude $\mu$~\citep{Holsclaw2010} are inferred from an observed photon flux
\begin{equation}
	\label{dL_z_def}
d_{L}(z)= (c/H_0)(1+z) D_L(z),
\end{equation}
\begin{equation}
	\label{mu_z_DL}
\mu(z)=5 \lg \left( d_{L}(z) \right)+ 25,
\end{equation}
where $d_L$ has units of $Mpc$, $H_0$ is the present Hubble constant and the dimensionless luminosity distance is
\begin{equation}
	\label{def_DL}
D_L(z)=\int_0^z \frac{dZ}{H(Z)/H_0},
\end{equation}
with $H(Z)$ being Hubble constant at redshift $Z$.}

At this point it is necessary to note that in addition to specifying the relevant equations of state, the full cosmological model also 
depends on four parameters. In the general case of three components (matter, radiation and dark energy), these parameters are the present 
Hubble constant $H_0$, and the present day total, radiation and dark energy density parameters, $\Omega_{tot0}$, $\Omega_{\gamma 0}$ and 
$\Omega_{\Lambda 0}$, respectively. These are in principle independent of any given dark energy EoS. $D_L$ is affected by the $\Omega$'s, 
while $H_0$ enters $\mu$ to give the dimensions of distance in Eq. (\ref{mu_z_DL}). Therefore, for every set of $\Omega$'s one requires a 
new solution to the differential equation, while different values of $H_0$ will just rescale the distance modulus.

We introduce the dark energy EoS as:
\begin{equation}
	\label{defDEEoS}
w_{\Lambda} \equiv \frac{p_{\Lambda}}{\epsilon_{\Lambda}},
\end{equation}
where $\epsilon_{\Lambda} = \Lambda/(8 \pi G_N)$.

Considering matter to be pressureless ($p_m=0$) allows to express the effective $\tilde{w}$ in terms of $w_{\Lambda}$ as:
\begin{eqnarray}
	\label{defDEEoS2tres}
&& \tilde{w}(x) = \frac{p_m+p_\Lambda+p_\gamma}{\tilde{\epsilon}}
= \nonumber \\ 
&& w_{\Lambda}(x) \frac{\Omega_{\Lambda}(x)}{\Omega_{tot}(x)}+\frac{1}{3} \frac{\Omega_{\gamma}(x)}{\Omega_{tot}(x)} =
w_{\Lambda}(x)+ \\
&& \frac{(1/3 - w_{\Lambda})\Omega_{\gamma 0} x^4-w_{\Lambda}(\Omega_{tot0}-\Omega_{\Lambda0}-\Omega_{\gamma 0})x^3}{f(x)+(\Omega_{tot0}-1)x^2},
\nonumber
\end{eqnarray}
{\color{black} where matter $\Omega_{m0}=\Omega_{tot0}-\Omega_{\Lambda 0}-\Omega_{\gamma 0}$, $x=1+z$, $f(x)=(H(z)/H_0)^2$ (as well as ratio of critical energy density to its recent value) so $f(1)=1$.}

We shall now neglect the energy contribution from radiation ($\Omega_{\gamma}=0$), valid for the present expansion epoch, and split 
the cosmological fluid into two components: pressureless matter ($w_m=0$) and dark energy ($\Lambda$). Thus,
\begin{equation}
\label{Omtotdos}
 \Omega_{tot} =\Omega_{\Lambda}+\Omega_{m}
\end{equation}
and therefore Eq. (\ref{defDEEoS2tres}) becomes
\begin{eqnarray}
	\label{defDEEoS2dos}
\tilde{w}(x) = \frac{0+p_\Lambda}{\epsilon}=w_{\Lambda}(x) \frac{\Omega_{\Lambda}(x)}{\Omega_{tot}(x)}= && \nonumber \\
w_{\Lambda}(x) \left(1-\frac{(\Omega_{tot0}-\Omega_{\Lambda0})x^3}{f(x)+(\Omega_{tot0}-1)x^2}\right).&&
\end{eqnarray}

We now introduce the assumption of a flat cosmology, amply justified by observations of the CMB anisotropy and power spectrum 
inferences from luminous red galaxies in the Sloan Digital Sky Survey (SDSS) e.g \citep{Tegmark2006}, giving $\Omega_{tot} = 1$. 
With the above, Eq. (\ref{defDEEoS2dos}) simplifies to
\begin{equation}
	\label{defDEEoS2}
\tilde{w}(x) = w_{\Lambda}(x) \left(1-(1- \Omega_{\Lambda0})\frac{x^3}{f_{flat}(x)}\right),
\end{equation}
{\color{black}where $f_{flat}(x)$ is the solution of evolution equation
\begin{equation}
	\label{eqfxflat}
f'_{flat}(x) \, x = 3 \, (1 + \tilde{w}(x)) \, f_{flat}(x) .
\end{equation}
}

Therefore, in the particular case of a flat universe we have only two scalar parameters to vary, $\Omega_{\Lambda0}$ and $H_0$, in addition to the dark energy equation of state, the function $w(z)$.

In general, the EoS $p(\epsilon)$ adiabatic parameter $w$ is defined as the ratio of the pressure $p$ to the energy density $\epsilon$ 
of any given component,
\begin{equation}
	\label{w_def}
	w \equiv \frac{p}{\epsilon}.
\end{equation}

Given that the only unknown EoS remaining in the model is that of the dark matter component, from this point onwards we sometimes drop 
the subscript from the dark energy EoS.

To generate a set of random $w(z)$ functions we begin with randomly choosing an initial point $w_0=w(z=0)$. As already mentioned, any 
EoS will be assumed to be constant at $w(z)=w_0$ for $z<0.01$. Next, we discretize the EoS space in the range of interest 
$-2 \le \lg(z) \le 0.2$, working in a  $w, \lg(z)$ plane. Note the use of $\lg$ for base ten logarithms, and of $\ln$ for natural logarithms. 
In this plane we generate curves through linear segments with random slope $\alpha_s$ and random length at every step $s$
\begin{equation} 
	\label{gen_tree}
	w_s(z)=w_{s-1}(z_s)+\alpha_s \left(\lg(z)-\lg(z_s)\right),
\end{equation}
where $z_s \leq z < z_{s+1}$. A class EoS's constant throughout the full redshift range of data are also trivially generated, 
and added to the full set of random walk curves treated.

The curves are then smoothed through the use of a Gaussian filter.
The coefficients of the filter are chosen to be
\begin{equation}
	c_i \equiv e^{-\frac{(i/l)^2}{\sqrt{2}}}/\sum_{j=-l}^{l}{c_j}.
	\label{gaus1}
\end{equation}
where $i=-l, -l+1, ..., l$, with the sum a normalisation. Hence, smoothing by the filter function $f(x)$ is done using $2l+1$ 
neighbouring points $x_i$ at every point $x_j$
\begin{equation}
	f_s(x_j) = \sum_{i=-l}^{l}{c_i \, f(x_{j+i})}.
	\label{gaus2}
\end{equation}
We choose nine neighbor points out of ten per segment, $l=4$ for our purpose. Examples of a few randomly generated EoS are shown 
in Fig. \ref{exmplDisEOS}.

\section{Bayesian $w(z)$ inferences from SNeIa, BAO and GRB constraints and tests of the method}

\begin{figure}[thb]
\vspace{-20pt}
\includegraphics[width=7.5cm]{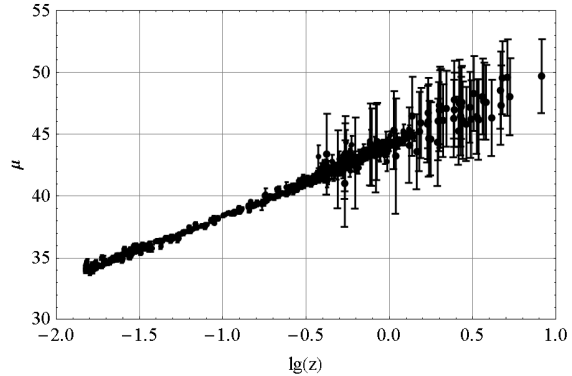}
\vspace{-20pt}
\caption{Luminosity distance modulus versus logarithm of redshift for SN-Ia and corresponding error bars, $(z_j, \mu_j \pm \Delta \mu_j)$. 
GRBs are inferred from the correlation assuming a flat $w=-1$ cosmology and stand out only from their larger error bars, no discontinuity 
is evident, implying a first order consistency of a $w=-1$ model out to very high redshift. The SN-Ia data were taken from the Union 2.1 
compendium \citep{SN_Union2p1}.}
\label{SN_Union2p1GRB}
\end{figure}

To begin with we consider data from a recent sample of SNeIa ''standard candles''. We take $N_{SN}=580$ SN events from the Union 2.1 
compendium \citep{SN_Union2p1}. As a first test, we treat our $N_{GRB}=54$ GRBs assuming the proposed correlation, and only here,
also a standard flat $w=-1$  cosmology (e.g. \citep{Capozziello2012, Capozziello2011, Izzo2009}), to obtain a first GRB $\mu,z$
catalogue.  Luminosity distance moduli $\mu$ are defined in 
Eq. (\ref{mu_z_DL}). Redshifts and $\mu$ values for both samples are shown in Fig. \ref{SN_Union2p1GRB}, where we see that despite the 
larger error bars associated to the GRB sample, no discontinuity is evident between the two data sets.  Having used the correlation 
coefficients inferred for the SNe Ia overlap region, where both data samples are clearly compatible, has yielded a consistent Hubble plot
extending out to $z=8.2$.
Notice the close to one order of magnitude extension in redshift range provided by the GRB sample, which in turn 
permits the possibility of tracing the cosmological model out to the extremely large redshift of 8.2. Another observational constraint 
come from BAO measurements, essentially a ''standard ruler'' at redshift $z_{BAO}=0.35$, characterized by the dimensionless parameter 
$A$ \citep{Tegmark2006, Adak2011}:
\begin{equation}
	\label{ang_BAO}
A = \sqrt{\Omega_{m0}}\left(\frac{D_L(z_{BAO})^2}{z_{BAO}^2 \sqrt{f(z_{BAO})}}\right)^{1/3},
\end{equation}
where $\Omega_{m0}$ is value of the present day matter density parameter and
\begin{equation}
	\label{A_BAO}
A = 0.469 \pm 0.017.
\end{equation}
Thus, $A_{BAO}=0.469$ and $\Delta A_{BAO}=0.017$ represent the central value and a $1\sigma$ confidence interval on the BAO observations. 
Data points for the SNeIa sample are magnitudes $\mu_j$ with error-bars $\Delta\mu_j$ at redshift $z_j$ (indexed by $j = 1 ... N_D$) with 
$N_D=N_{SN}$ being total number of available events. 

To every proposed model (identified by the index $n$), i.e. a set of ($H_{0}, \Omega_{\Lambda0}$) values and a corresponding generated $w(z)$ 
curve within the redshift interval chosen, a probability related to $\chi^2$ is assigned as a measure of how well each model represents the 
various data sets, the SNeIa, and the BAO constraint. Assuming each data set as statistically independent, this $\chi^2$ takes the form: 
\begin{equation}
 \chi_{n}^2=\chi_{n,SN}^2+\chi_{n,BAO}^2,
 	\label{chi2n}
\end{equation}
where we define
\begin{equation}
 \chi_{n,SN}^2=\frac{1}{N_{D}}\sum_{j=1}^{N_{D}}(\mu_j-\mu_n(z_j))^2/{\Delta\mu_j}^2,
 	\label{chi2j}
\end{equation}	
\begin{equation}
 \chi_{n,BAO}^2=(A_{BAO}-A_n)^2/\Delta A_{BAO}^2.
 	\label{chi2bao}
\end{equation}
Note that although the BAO observation has the same statistical weight as one SN, the fact that this is very localized in redshift space, 
and the tight confidence interval for $\Delta A_{BAO}=0.017$ make it a valuable constraint, effectively narrowing substantially the range of 
$\Omega_{\Lambda0}$ and allowed $w(z)$ solutions at the BAO redshift.
\begin{figure*}[htb]
\vspace{-80pt}
\includegraphics[width=16cm]{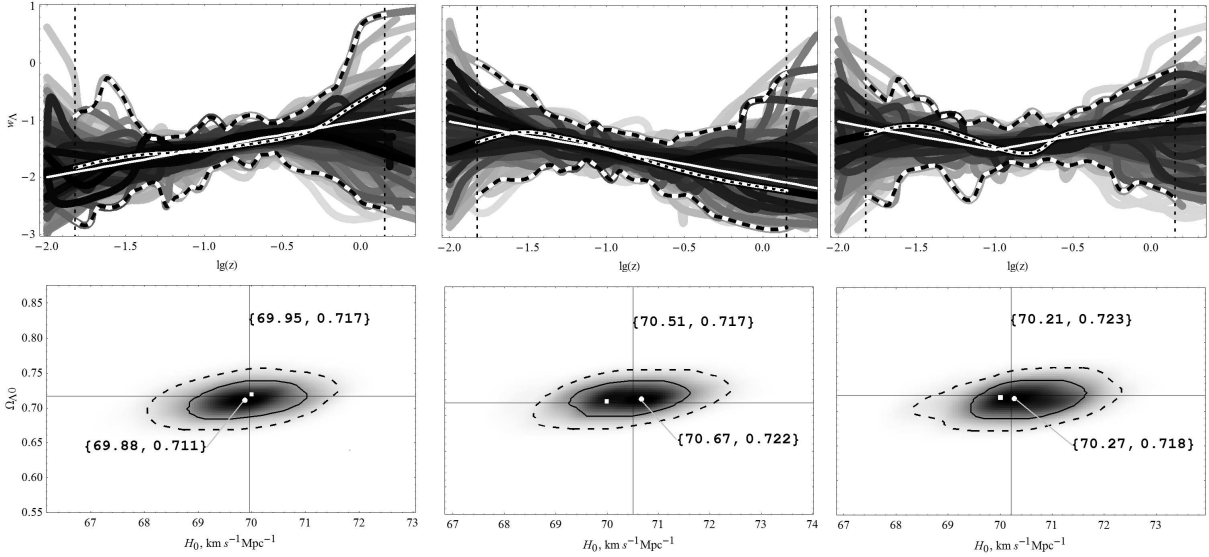}
\vspace{-80pt}
\caption{Reconstruction of test EoS using simulated SN-Ia and BAO data. Three cases are shown by the straight white lines (from left 
to right): rising, falling and mixed. Corresponding parameter maps are shown below. Crossing lines are input values, squares give the 
maximum probability cases and solid dots the maximum after marginalization.}
\label{tests}
\end{figure*}
%

Since $\mu$ scales with $H_0$ in the logarithm, a further simplification is possible.
Let us define 
\begin{equation}
 \mu_H \equiv 5 \lg(c/H_0),
 	\label{muH}
\end{equation}
\begin{equation}
 \widetilde{\mu}_n \equiv 5 \lg \left((1+z)D_L(z,n) \right)+25,
 	\label{muDOm}
\end{equation}
and introduce the following parameters
\begin{equation}
 \alpha=\sum_{j}^{N_D}\frac{\mu_j}{\Delta \mu_j^2},
 	\label{almu}
\end{equation}
\begin{equation}
 \beta_n=\sum_{j}^{N_D}\frac{\widetilde{\mu}_n(z_j)}{\Delta \mu_j^2},
 	\label{betmu}
\end{equation}
\begin{equation}
 \gamma=\sum_{j}^{N_D}\frac{1}{\Delta \mu_j^2}.
 	\label{gammu}
\end{equation}
And use them to define
\begin{equation}
 \mu_{D,n}=\frac{\alpha-\beta_n}{\gamma},
 	\label{muDn}
\end{equation}
\begin{equation}
 \Delta \mu=1/\sqrt{\gamma};
 	\label{Dmu}
\end{equation}
therefore we can split as
\begin{equation}
 \chi_{n,SN}^2=\chi_{n,\Omega}^2+\chi_{n,H}^2/N_D,
 	\label{chi2s}
\end{equation}
where
\begin{equation}
 \chi_{n,H}^2=\frac{(\mu_H-\mu_{D,n})^2}{\Delta \mu^2},
 	\label{chi2Om}
\end{equation}
and
\begin{equation}
 \chi_{n,\Omega}^2=\frac{1}{N_D}\sum_{j}^{N_D}\frac{(\widetilde{\mu}_n(z_j)-\mu_j)^2}{\Delta \mu_j^2}-\frac{\mu^2_{D,n}}{N_D \Delta \mu^2}.
 	\label{chi2H}
\end{equation}
Now
\begin{equation}
H_n=c \, 10^{-\mu_{D,n}/5},
\label{Hndef}
\end{equation} 
gives the value of $H_0$ which minimizes the $\chi^2$  for a given $w(z)$ and $\Omega_{\Lambda0}$.

Although complex probability density functions could be introduced at this point to take advantage of the Bayesian approach, for this  
introductory work (and lacking any evidence to the contrary) every data point $j$ (with measured redshift $z_j$) is treated as a normal 
probability density function $f_j$ with mean value $y_j$ and standard deviation $\Delta y_j$, taken from observations,
\begin{equation}
	f_j(y)=\frac{e^{-(y-y_j)^2/{2(\Delta y_j)}^2}}{\sqrt{2 \pi} \Delta y_j},
	\label{norm_distr}
\end{equation}
therefore Bayesian analysis turns into maximum likelihood with minimization of $\chi^2$. A particular full cosmological model will now yield 
a fixed curve $\mu_n(z; H_0, \Omega_{\Lambda0}, w(z))$, with a likelihood assignment at redshift $z_j$ given by:
\begin{equation}
	p(D_j|H_{yn})=f_j\left(\mu_n(z_j; H_0, \Omega_{\Lambda0},w(z))\right),
	\label{pDHj}
\end{equation}
where $p$ is thus the relative probability of this model with respect to another one. Sequential application on all data points of 
Eq. (\ref{Bayesian_seq}) results in the total likelihood given by the product
\begin{equation}
	p(SN|H_{yn})=\prod_j^{N_D}{f_j\left(\mu_n(z_j; H_0, \Omega_{\Lambda0})\right)},
	\label{seq_pDH}
\end{equation}
and for BAO
\begin{equation}
	p(BAO|H_{yn})=\frac{e^{-(A_{BAO}-A_n)^2/{2(\Delta A_{BAO})}^2}}{\sqrt{2 \pi} \Delta A_{BAO}}.
	\label{norm_distrBAO}
\end{equation}

We next apply Eq. (\ref{seq_pDH}) and  Eq. (\ref{norm_distrBAO}) to all generated curves. The result is that every hypothesis 
$H_y$(a $w(z)$ curve with particular $H_0$ and $\Omega_{\Lambda 0}$ parameters values) acquires probability 
\begin{eqnarray}
		p(H_{yn}|D) && \propto p(BAO|H_{yn}) p(SN|H_{yn}) \nonumber \\
							&& \propto e^{-\chi^2_{n,BAO}} e^{- {N_D} \, \chi^2_{n,SN}}\\
							&& =  e^{-\chi^2_{n,BAO}} e^{- {N_D} \, \chi^2_{n,\Omega}} e^{\frac{(5 \lg(H_n/H_0))^2}{2 \Delta \mu^2}}. \nonumber
	\label{seq_pHDtot}
\end{eqnarray}
Notice that for the particular case of a normal error distribution function, the dependence of the probability on the Hubble 
constant can be calculated analytically, saving computation time.

Next we choose the curve with maximum probability, $p_{max}$, and order the rest in descending order by value of relative probability 
$p \to p/p_{max}$, which is also used for scaling in grey (black corresponding to the maximum). This helps to visualise the regions where 
curves with the highest probability go through, and therefore where the real dark energy EoS is most probably located.
The size and shape of the region depends on the number of data points, their quality (errors) and distribution
(in redshift bins).

Finally, we must derive confidence intervals around the optimal solution, for this we shall construct a band of EoS's encompassing the 
most likely curves and bounding the variations allowed by the data. We begin by taking the maximum probability EoS $w_{max}(z)$ and 
construct boundary curves which accumulate $68.2 \% $ of the total probability $P_{t}=\sum_n{p_n}$, following in the sum a descending order 
in $n$ we have constructed:
\begin{eqnarray}
	w_{+}(z)=\max(w_{n}(z) | \sum_{ordered}{p_n}/P_{t} \le 0.682), \nonumber \\
	w_{-}(z)=\min(w_{n}(z) | \sum_{ordered}{p_n}/P_{t} \le 0.682).
	\label{boundsw}
\end{eqnarray}
To construct a relative probability map for parameters $H_0$ and $\Omega_{\Lambda0}$ we marginalize as
\begin{equation}
	p(H_0,\Omega_{\Lambda 0})=\sum_{n} p(H_0,\Omega_{\Lambda 0},EoS_n).
	\label{pHOmMAP}
\end{equation}
We locate the maximum probability model as well as the maximum point 
$({H_0,\Omega_{\Lambda0}})_{m}$ after marginalization. In order to estimate $1-\sigma$
and $2-\sigma$ contours we locate the contours where probability accumulates $68\%$ and $95\%$ of the total, respectively.

Since every model of particular $w(z)$ includes set of initial cosmological parameters we
marginalize over them to get probability band for EoS only
\begin{equation}
	p(EoS_n)=\sum_{\Omega_{\Lambda0}} \int_{H_0} p(H_0,\Omega_{\Lambda 0},EoS_n).
	\label{margp}
\end{equation}

The code begins by generating $5000$ random $w(z)$ curves as described in section (3), together with $2000$ constant $w$ curves with 
values constrained to the range $-4 \le w \le 2$ including one curve for a classical cosmological constant EoS, $w(z)=-1$, $7000$ curves 
in total. The length of every line segment is randomly distributed between $0.1$ and $0.5$, and the angle of every segment is randomly 
chosen between $-89^o$ and $+89^o$, in the $\lg(z)$ vs $w$ plane.

The two scalar parameters of the flat universe run through $50$ discrete values in range $\Omega_{\Lambda0}=0 ... 1$ and $50$ in 
$\Omega_{\Lambda0}=0.5 ... 0.9$. To avoid large quantities of low probability cases we prune the initial random EoS's to those within 
$N_\sigma=3$, estimating $n \sigma$ cuts through
\begin{equation}
	\label{defNsigma}
	\ln(p) > \ln (p_{max})-N_{\sigma}^2/2,
\end{equation}
based on a correspondence with the maximum likelihood method.


In this section we shall perform a number of tests designed to asses the robustness of the methodology presented, firstly in a case where the
answer is actually known, i.e., through the use of synthetic data samples constructed using a particular cosmological model, an
input $w(z)$ assumption. In this way, the results and their corresponding confidence intervals can be compared 
to the input model, for a variety of controlled possibilities.

Redshifts and normal error amplitudes are taken from the original SNeIa data set. The results of the test are shown in Fig. \ref{tests}. 
We see the recovered EoS, given by the white curves, always very close to the input value, shown by the thin white lines. The region between 
dashed curves gives the estimated $1\sigma$ (cumulative $68\%$) confidence band. We see that even though the method intrinsically samples 
non-linear $w(z)$, the data drive the solution to a linear value, coinciding with the input one. Notice also that the final inferred 
$H_{0}, \Omega_{\Lambda0}$ values appear very close to the input ones, see bottom row in Fig. \ref{tests}. 


As will be seen from SNeIa and BAO low redshift ($z<1$) EoS is consistent with $w(z)=-1$
as well as $H_{0}=70$ ${\rm km} \, {\rm s}^{-1} {\rm Mpc}^{-1}$ and $\Omega_{\Lambda 0}=0.72$.
So we use that model and construct random tree of EoS above for higher redshift $z>1$
to study constraining power of our GRB sample. Data is analyzed by forming a distance ladder.
\begin{figure}[htb]
\vspace{-20pt}
	\includegraphics[width=7.5cm]{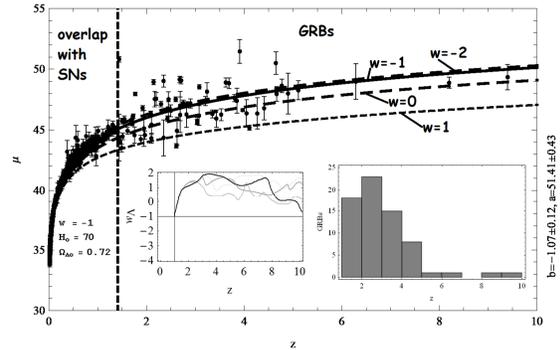}
\vspace{-20pt}
\caption{Distance ladder. GRBs in the SNeIa overlap redshift range, where cosmology is well constrained, are used to calculate the GRB 
intrinsic correlation coefficients. This correlation is then used to calculate the luminosity distance for high redshift GRBs from their 
X-ray afterglow luminosity curves. Standard constant $w$ solutions are shown for reference. Vertical dashed line marks farthest SN-Ia 
event. Inset to the right shows a histogram of our GRB sample distribution in redshift. Inset to the left shows resulting most probable EoS, 
together with a small sample of models probed, confidence intervals are so large, that only extreme variations with respect to $w=-1$
can be excluded.}
\label{GRBladder}
\end{figure}
As seen from Fig. \ref{GRBladder},
once correlation coefficients are calculated from overlap region ($33$ GRBs) with SNeIa and
then used to get luminosity distance for higher $z$ $68$ GRBs, they are quite scattered
and hard to fit well ($\chi^2$ is large) with randomly generated smooth $w(z)$ curve (see the inset). Also
major number of GRBs are detected in $z<4$ as the other inset with histogram shows. 
Usage of correlation for calculation of luminosity distance of GRBs beyond SNeIa introduces
correlation between data points of the different GRB events. Once correlation matrix $C_{ij}$ is
known the Eq. (\ref{chi2j}) has to be modified into
\begin{equation}
 \chi_{n,GRB}^2=\frac{1}{N_{D}}\sum_{i,j}^{N_{D}}(\mu_i-\mu_n(z_i))C_{ij}(\mu_j-\mu_n(z_j)),
 	\label{chi2jcorr}
\end{equation}	
where now $N_D=N_{GRB}=68$.
What one can do is to impose $\mu$ to lie on standard case of $w=-1$ and study constraining power
of correlated errors as well as effect of data distribution in redshift bins.
\begin{figure*}[htb]
\vspace{-30pt}
\includegraphics[width=16cm]{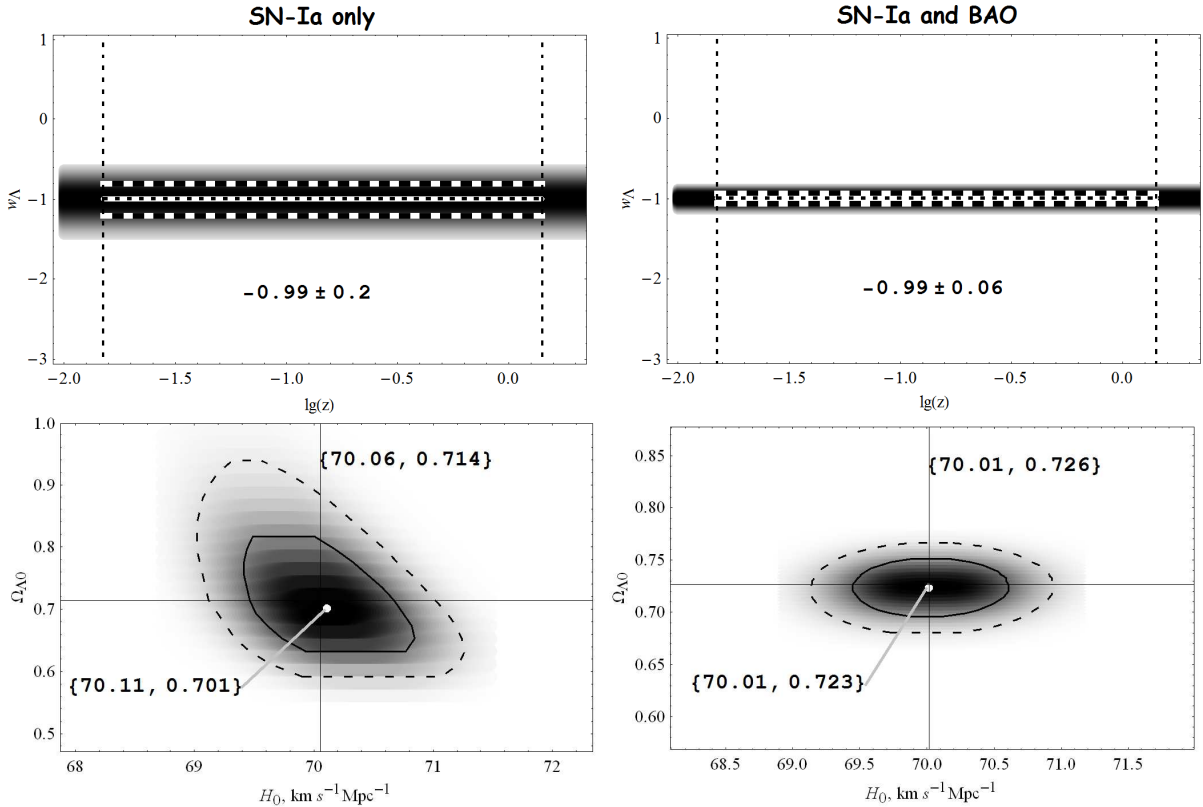}
\vspace{-30pt}
\caption{EoS's shaded according to relative probability after applying our Bayesian analysis to the class of constant curves and using 
only SNeIa data (left column) and with the addition of BAO constraint (right column). The white central lines are the maximum probability 
$w_{max}(z)$, whereas the dashed lines give the confidence intervals (cumulative $68\%$). Vertical dashed lines mark the data range. 
The bottom row gives corresponding probability maps for the scalar cosmological parameters of the fit. White dots mark the marginalised 
maxima and crossed lines show the maximum probability cases.}
\label{finalSNandBAOconstants}
\end{figure*}

\section{Results}

To begin with, we analyze only $2000$ constant EoS's evenly spaced between $-4 \leq w_{\Lambda} \leq 2$. Starting only with the SNeIa data 
sample, we will study the effect of adding the BAO constraint later. Results are shown in the left column of 
Fig. \ref{finalSNandBAOconstants}. 
We see a very well defined solution consistent with a cosmological constant, and having a narrow confidence interval, $w=-0.99 \pm 0.2$.
Adding the BAO constraint, right panel, does not change the central result, but considerably reduces the confidence interval, resulting
now in $w=-0.99 \pm 0.06$. The inferred scalar parameters appear in the lower row, where again, we see the inclusion of the BAO
constraint significantly tightening the confidence region, particularly in the case of the the inferred dark energy density parameter, 
yielding the very precise estimate of $\Omega_{\Lambda0}=0.723 \pm 0.025$. in the lower row we see in general the very high 
resolution with which the method infers the present day cosmological parameters $H_{0}, \Omega_{\Lambda0}$, once all the data 
are taken into account.
\begin{figure}[htb]
\vspace{-15pt}
\includegraphics[width=7.5cm]{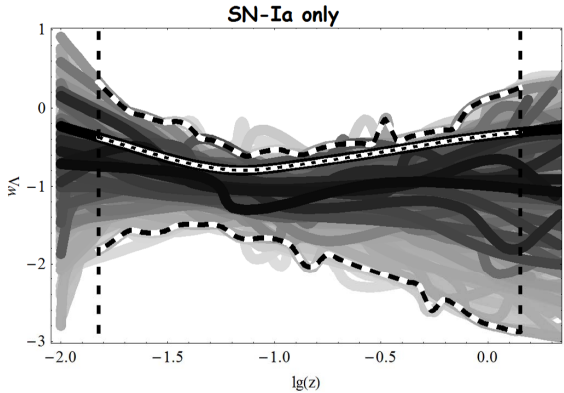}
\vspace{-15pt}
\caption{Resulting relative probability (shaded in gray) for $w(z)$ when only SNeIa data is analysed.
Marginalisation of Eq. (\ref{margp}) is applied. Maximum probability (white) and cumulative $68\%$ boundaries
(striped) are shown. Vertical dashed lines show SNeIa data range.}
\label{finalSNeos}
\end{figure}
\begin{figure}[htb]
\vspace{-15pt}
\includegraphics[width=7.5cm]{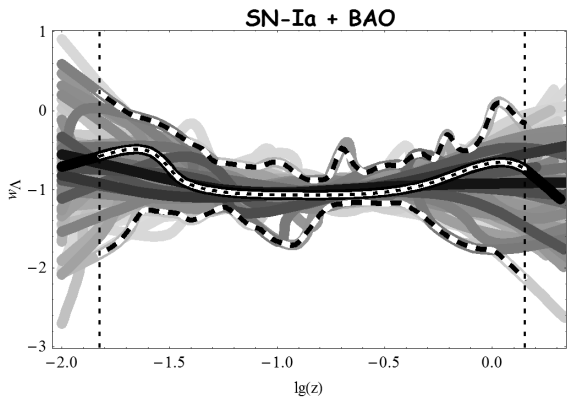}
\vspace{-15pt}
\caption{Same as Fig. \ref{finalSNeos} but after the BAO constrain is included in the analysis.}
\label{finalSNBAOeos}
\end{figure}

If we now allow the data to select the $w(z)$ model which best reproduces the luminosity distance and redshifts of the SNeIa sample and 
the observed BAO constraint, we obtain results shown in Fig. \ref{finalSNeos} and Fig. \ref{finalSNBAOeos}, for the
inferred $w(z)$ dark energy EoS. Every EoS is shaded in gray according to its relative probability $p/p_{max}$, where $1$ is black and 
$0$ is white. It is seen that when using SNeIa data only, at higher redshift ( $\lg(z)>0$) curves explore a wide range of possibilities, 
not being restricted by any data. Further addition of the BAO constraint significantly tightens the confidence band of the allowed $w(z)$ 
curves. Our principal result is now given in terms of the maximum probability dark energy $w(z)$ and the corresponding confidence interval 
bands covered by the distribution of lower probability equations of state.  We see our inference lying extremely close to the classical 
cosmological constant of $w(z)=-1$, once the BAO constraint is included. In the figures, the maximum probability curves are given by the 
white dashed line, while a representative sample of the other $w(z)$ curves explored is given in shades of gray.
\begin{figure}[htb]
\vspace{-15pt}
\includegraphics[width=7.5cm]{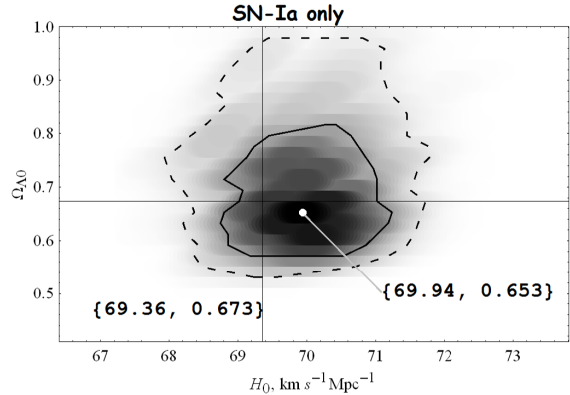}
\vspace{-15pt}
\caption{Complimentary to Fig. \ref{finalSNeos}, probability map of inferred cosmological parameters.
Marginalisation of Eq. (\ref{pHOmMAP}) is applied. Crossing lines give the maximum probability 
model before marginalisation and solid dot shows the probability maximum after marginalisation. Contours
enclose $68\%$ and $95\%$ of cumulative probability.}
\label{finalSNmap}
\end{figure}
\begin{figure}[htb]
\vspace{-15pt}
\includegraphics[width=7.5cm]{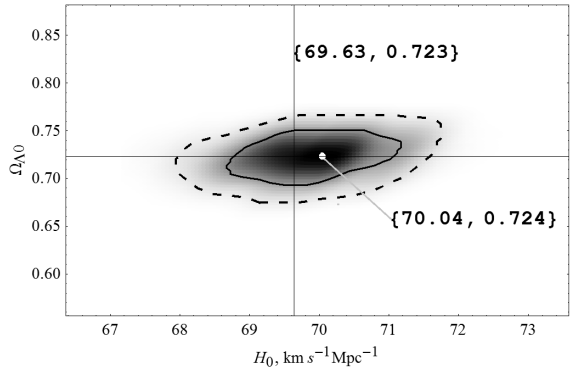}
\vspace{-15pt}
\caption{Same as Fig. \ref{finalSNmap} but after the BAO constrain is included in the analysis.}
\label{finalSNBAOmap}
\end{figure}

\begin{figure*}[htb]
\vspace{-110pt}
	\includegraphics[width=16cm]{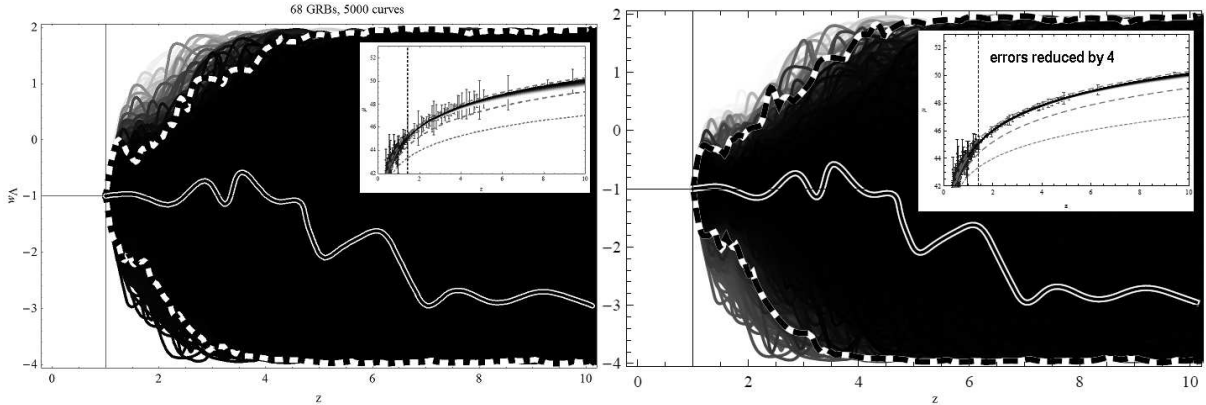}
\vspace{-110pt}
\caption{Tree of $w(z>1)$ curves inferred from synthetic GRB samples {\color{black}constructed for $w(z)=-1$ cosmologies}, showing
to what extent correlated GRB errors constrain EoS at high ($z>1$) redshifts. Using GRB errors taken from actual data,
left panel, and GRB errors reduced by a factor of $4$, right panel.}
\label{centeredDATAtwo}
\end{figure*}

Adding GRB data will in principle narrow the parameter space explored by the method, asides from yielding 
a one order of magnitude extension in the redshift range probed. However, given the small number of points available over a large redshift
range, and the large error bars associated to these data, the power to constrain the high redshift $w(z)$ form, is presently almost
absent. This is seen in Fig. \ref{centeredDATAtwo}, where the method is fed a synthetic GRB sample having the same redshift distribution
and error bars as the real data, but sampled directly from a $w=-1$ universe, left panel. We see only very extreme $w(z)$ variations
are excluded, and within the confidence region, an almost flat (constant color) likelihood surface. We can explore what added constraining
power a better future sample might yield, e.g. repeating the experiment including a reduction in the errors by a factor of 4, right panel.
This, even without considering any expected increase in the number of GRB events, would already permit much more interesting
high redshift dark energy constraints. We note \citet{Dainotti2010ApJ722L215D} showed that such a decrease in the errors of GRB data
is certainly feasible.

There are several reasons for the low $w(z)$ constraining power of high 
redshift events: the expansion of the universe becomes dominated by matter, and hence the sensitivity of distance luminosity to the dark 
component is quite small. Also, the low number of events in the SNe Ia overlap region implies large error bars on the inferred 
GRB correlation coefficients, while the large error bars, despite their correlations, for high $z$ GRBs produce a very flat probability 
(seen as the uniform black shading in fig\ref{centeredDATAtwo}, left panel) distribution for the different EoSs tested. 
Having said that, there is much expectation for the $1<z<4$ region, once the GRB sample is increased and its quality is improved.

The corresponding scalar parameter $H_0$, $\Omega_{\Lambda0}$ inferences for the method applied to the SNa Ia data sample and the BAO 
constraint, allowing for free $w(z)$ variations, are shown in Fig. \ref{finalSNmap} and Fig. \ref{finalSNBAOmap}.

The BAO constraint was sequentially added to the data set considered. In all cases above, the two contours give $1\sigma$ and $2\sigma$ 
confidence regions. Notice that in all our results, once the data samples are considered, in terms of our inferred $w(z)$ and $H_{0}$ and 
$\Omega_{\Lambda0}$ values, probability distributions peak far from the borders of the parameter region sampled; in all cases our confidence 
intervals are closed, showing that the inferred model is not only the best fit to the data amongst the many sampled, but also that it is a 
fair representation of the data considered, as verified by the corresponding chi squared values obtained.

We next check the look-back time for the different cosmological models tested, which are consistent with a comparison to a lower age limit 
coming from globular cluster ages of 13 Gyr, \citep{Mackey2002tn,Mackey2002un,Mackey2002jm}. The maximum probability curves yield ages
always above 13 Gyr, and hence are fully consistent with this extra condition. 

Finally, we test the robustness of our results to the details of the $w(z)$ sampling method used. We now implement 
an independent curve generating algorithm based on a Markov chain formalism. Similarly to what was done previously, a chain dark 
energy EoS is formed from piecewise curves with end points chosen at redshifts $\lg(z_0)=-2$ and $\lg(z_{N_{nodes}+1})=1$, and $N_{nodes}$ 
in between.

\begin{figure}[htb]
\vspace{-10pt}
\includegraphics[width=7.5cm]{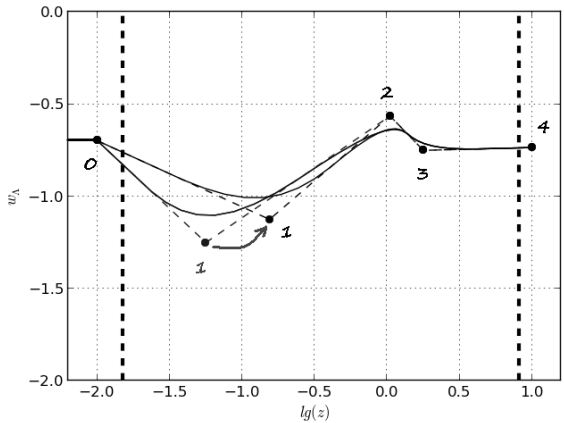}
\vspace{-10pt}
\caption{Example of a Markov chain step in the sampled equations of state. The case of $N_{node}=3$. Dashed lines represent
curves generated by randomly moving random node. Superimposed solid line is the result of applying the smoothing Gaussian filter. Vertical dashed lines mark data range.}
\label{markovStep3nodesEOS}
\end{figure}

At each step we slightly modify the curve by randomly choosing a node (including the endpoints) and randomly moving it in the EoS plane, 
where the endpoints are only allowed to move vertically, stepping in $w_{\Lambda}$. The resulting curve is then smoothed using the same 
Gaussian filter used previously both in the $\lg(z)$ and $w_{\Lambda}$ directions, a sample of such curves is shown in 
Fig. \ref{markovStep3nodesEOS}. This final curves are then passed to the same likelihood calculating procedure described previously,
using both the SNeIa data and the BAO constraint. To these we add a synthetic sample of high redshift GRBs having the same redshift 
distribution and errors as the real data, and generated for a $w(z)=-1$ universe, in order to asses the constraining power of our current 
GRB sample.

If the Bayesian probability of the modified curve is increased relative to the previous one, then this new curve is picked for new adjustments 
at the next step, otherwise a random number between 0 and 1 is selected, with the new EoS rejected if the following condition holds:
\begin{equation}
	\label{dice}
	r_{die}[0...1] > p_{modified}/p_{initial},
\end{equation}
where $p$ is the Bayesian probability of the corresponding curve.

In this way we explore the EoS space close to the most probable cases and spend less time on less probable ones. After a significant number 
of steps the result is independent of the particular EoS curve chosen to begin with.
\begin{figure}[htb]
\vspace{-10pt}
\includegraphics[width=7.5cm]{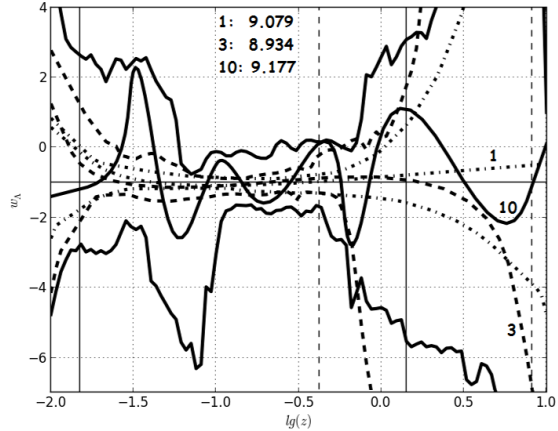}
\vspace{-10pt}
\caption{Maximum relative probability $w(z)$ given by the Markov chain method and cumulative $68.2\%$ 
confidence intervals. Three cases are shown: $N_{node}=1$ (dot-dashed), $3$(dashed) and $10$(solid). Vertical lines mark data ranges. Numbers correspond to natural logarithm of the maximum probability (not normalized).}
\label{eos1and3and10nodesBAOSNGRBmore}
\end{figure}
\begin{figure}[htb]
\vspace{-10pt}
\includegraphics[width=7.5cm]{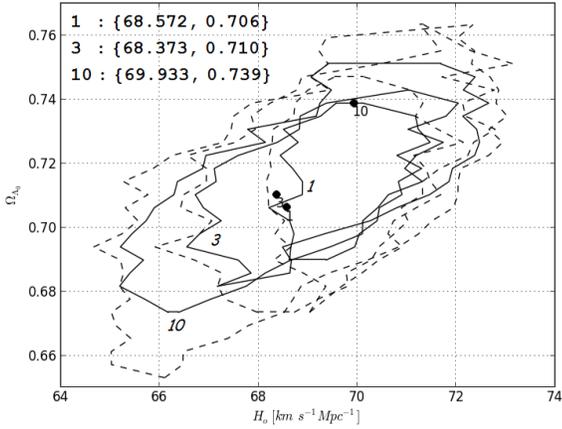}
\vspace{-10pt}
\caption{Recovered parameters using the Markov chain method with cumulative $68.2\%$(solid) and $95.4\%$(dashed) contours.
Three cases are shown: $N_{node}=1$, $3$ and $10$. Circles mark location of the maximum probability.}
\label{map1and3and10nodes936BAOSNGRB}
\end{figure}

We have run the alternative Markov chain method with over $10,000$ steps for three values of $N_{node}=1$, $3$ and $10$ and results 
are shown in  Fig. \ref{eos1and3and10nodesBAOSNGRBmore} and \ref{map1and3and10nodes936BAOSNGRB}. For one node we also tested for 
convergence using different initial conditions, which indeed were seen to quickly converge to consistent solutions. Increasing the 
number of nodes naturally increases the sensitivity of the method to redshift variations in $w(z)$, but also results in broader 
confidence intervals. Within the low redshift range, the final maximum probability EoS's and their confidence bands are very similar 
to what was obtained previously for each of the three cases considered in terms of the sequential data sets introduced, and remain 
consistent with the cosmological constant case of $w_{\Lambda}=-1$. The same applies to the probability contours in the $H_{0}, 
\Omega_{\Lambda 0}$ plane, hence proving the method to be fully robust with respect to the random $w(z)$ generation procedure and 
the subsequent exploration of the EoS plane; it is the data through the full Bayesian probability assignment which drive the inference. 
Regarding the high redshift $w(z)$ constraints, we see the results being consistent with the input $w=-1$ model, but having associated
confidence intervals which flare beyond the $z=1.4$ region.

\section{Conclusions}

We have developed and tested a sequential Bayesian analysis method tuned to the non-parametric inference of
the evolution of the dark energy equation of state.

Complementing SNeIa samples and BAO constraints with new GRB cosmological distance estimators will allow an order of magnitude extension
in the redshift range over which cosmological and dark energy physics can be traced, once error bars can be somewhat reduced, and the sample
extended. We obtain results consistent with
a cosmological constant $\Lambda CDM$ model, i.e. $w(z)=-1$ not requiring any redshift evolution out to $z=1.4$, although the 
confidence intervals obtained do allow for small variations throughout the redshift range sampled of $\pm 0.3$ at most.

Construction of a distance ladder and use of the full correlation matrix has been performed on a subclass of GRBs.
Analysis of our GRB sample and extrapolation of the local $w(z)$ EoS to high redshifts emphasizes the need of increasing the number
of data points (at least for $z<4$ and in the overlap region with SNeIa data) and to improve their quality, it is sufficient to reduce the 
error bars by a factor of $4$. Let us note here that such a subsample with low error bars is indeed realistic. In fact, in 
\citet{Dainotti2010ApJ722L215D} we demonstrated that the error bars of the GRB observables can be reduced by factors of 10 and more.

The method simultaneously yields optimal maximum likelihood inferences for the scalar parameters of the cosmological
model, considering the most general case of possible $w(z)$: $H_{0}=70.04 \pm 1$ ${\rm km} \, {\rm s}^{-1} {\rm Mpc}^{-1}$ and 
$\Omega_{\Lambda 0}=0.724 \pm 0.03$.

This result has a twofold interpretation: from one side our approach can not select amongst similar concurrent dark 
energy models at very high redshift. This could mean that we need further indicators and a wider redshift sample. On the 
other hand, this result could be read as a first indication of the fact that  the cosmological constant problem persists also 
at very high redshift. At a fundamental level, this feature would have severe consequences due to the difficulty in connecting the very 
tiny value of the observed cosmological constant with the vacuum state of the gravitational field at cosmological scales.

\section{Acknowledgments}

The authors acknowledge the careful reading of our first version by an anonymous referee, whose valuable input was important
in reaching a more compete presentation of our work. 
SP would like to acknowledge US National Science Foundation under grant $NSF-PHY-1205019$ and Mexican Conacyt grant CB-2009/132400. XH 
acknowledges financial assistance from UNAM DGAPA grant IN103011. M.G.D. acknowledges support 
from the Boncompagni-Ludovisi Scholarship and the Polish MNiSW through grant N N203 579840; and the initial support from the Fulbright 
and the {\it L'Oreal Italia per le donne e la Scienza} Scholarships. SC acknowledges the support of INFN, {\it Iniziative specifiche} 
NA12 and OG51.

\bibliographystyle{apj}
\bibliography{refsV33}

\end{document}